# Deep learning based inverse method for layout design

Yujie Zhang, Wenjing Ye[*]
Department of Mechanical and Aerospace Engineering
Hong Kong University of Science and Technology
*Corresponding author: mewye@ust.hk

**Abstract**

Layout design with complex constraints is a challenging problem to solve due to the non-uniqueness of the solution and the difficulties in incorporating the constraints into the conventional optimization-based methods. In this paper, we propose a design method based on the recently developed machine learning technique, Variational Autoencoder (VAE). We utilize the learning capability of the VAE to learn the constraints and the generative capability of the VAE to generate design candidates that automatically satisfy all the constraints. As such, no constraints need to be imposed during the design stage. In addition, we show that the VAE network is also capable of learning the underlying physics of the design problem, leading to an efficient design tool that does not need any physical simulation once the network is constructed. We demonstrated the performance of the method on two cases: inverse design of surface diffusion induced morphology change and mask design for optical microlithography.

## 1. Introduction

Inverse problems are encountered in many fields of engineering and science. In a typical inverse problem, parameters or layouts are sought in order to achieve certain system outcomes. Examples of inverse problems can be found in material design [1,2], structure optimization [3-6], determination of radiative properties of the medium [7,8], model parameter estimation [9,10], and image synthesis [11,12]. Most existing methods for solving inverse problems are based on optimization techniques, in which the desired system performance is casted as the objective function and the unknown parameters/layout are the design variables. Among the different types of design problems, layout designs are particularly challenging because of the difficulty in identifying a set of suitable design variables that define the layout. Topological optimization (TO) offers an attractive approach for solving this type of problem. Currently the most popular TO method is the density-based method in which the design domain is discretized into pixels/voxels and the optimal layout/structure is found by determining the material density of each pixel/voxel based on a chosen optimization strategy [13,14]. An alternative and powerful TO method is the level-set based method, which utilizes level-set functions to implicitly define the boundaries of the layout. The optimal layout is found by controlling the motion of the level-set functions according to the underlying physics and the optimization strategy [15-17]. In both methods, gradient-based optimization methods are commonly employed. Hence sensitivity calculation of the objective function with respect to the design variables is required. For dynamic problems, sensitivities could be difficult to obtain due to the high computational cost and the large memory consumption required to store all intermediate solutions [18-20]. In addition, not all the problems have differentiable objective functions, and thus non-gradient based approaches, such as genetic evolutional structure optimization, must be employed. However, these methods are known for their inefficiency because of the need for a large number of forward calculations. Another key challenge in topological optimization techniques is the control of the shape of the optimal layout. Shape control is necessary due to, for example, the need for including a few explicit shapes in the design, manufacturing constraints such as minimum feature size and manufacturing cost, aesthetic consideration and connection/installation requirements. These constraints are often cannot be formulated analytically in terms of design variables and the optimization parameters, and thus are difficult to be incorporated in the optimization process particularly when the density-based methods are employed. Recent work on overcoming this challenge includes the development of feature-driven, level-set function based TO methods to incorporate features into the freeform design domain [21,22], and the density-based TO method for imposing minimum and maximum length scales [23].

In recent years, machine learning approaches have shown remarkable success in a variety of application areas such as image recognition [24], natural language processing [25], PDE solver [26,27], quantum mechanics [28] and multiscale modeling of materials [29,30]. One major advantage of many machine-learning techniques is their ability of learning the hidden relationship of data, which is otherwise difficult to be revealed and modeled. Variational Auto-encoders (VAEs) are recently developed probabilistic models for learning complicated distributions of datasets [31,32]. The main function of a VAE is to provide a tractable and efficient mapping between latent variables, which follow a prescribed distribution, and the given dataset. By sampling the latent variables from the known distribution, this mapping can be used to generate new datapoints that share the same salient features as those in the given set. The applications of VAEs have been mostly in image processing [31,33-35], which was the targeted application initially. Since last year, innovative applications of VAEs in physical systems such as phase and phase transition identification, learning hard quantum distribution have emerged [36-38], demonstrating the great potential of VAEs in a variety of application areas. Inverse design with constrains is a problem of searching an optimal solution in a design space confined by constraints. To a large extent, one can view this design space as a specific one associated with a certain but unknown distribution. It is then possible to construct a VAE model to learn the unknown distribution by providing a dataset satisfying all the constraints. Once the mapping is established, the capability of the VAE of generating new datapoints that automatically satisfy the constraints would allow us to conduct the design without imposing any constraints since the search is within the confined space identified by the VAE. Apart from that, VAEs are built on top of neural networks, which are powerful function approximators. Thus the physical relationship between the inputs and the objective function can also be learnt by VAEs, leading to a design method that does not require any physical sensitivity calculations. Another potential benefit of the VAE is that once the model is constructed for a certain type of physical system, it can be used repeatedly to conduct designs with different objective function values and/or physical parameters without any physical simulation. There has no prior work on the applications of VAEs in solving inverse problems and thus the possible benefits of VAEs in this area have not been explored.

In this paper, we propose a VAE-based design method for inverse design of layout/structures. Of particular interest are problems with shape constraints. We construct the VAE model with neural networks and train the model using a dataset in which each datapoint consisting of two parts: the initial layout satisfying the shape constrains and the corresponding outcome that could be, for example, an image or a value. This allows the VAE to learn not only the shape constraints, but also the physical correlation between the initial layout and the outcome. Such a feature is particularly useful for problems in which the physical correlation is difficult or very costly to model. The final design is found by finding the optimal latent variables that generate the desired outcome using the decoding part of the VAE model, that is the generative feature of the VAE. Two very different problems namely mask design for optical microlithography and the inverse design for surface diffusion induced morphology change are selected to demonstrate the performance of the proposed method.

## 2. Neural network and inverse design

A standard VAE model contains an encoder and a decoder as shown schematically in **Fig. 1**. For each input datapoint, the encoder finds a corresponding latent variable from a prescribed distribution, typically the Gaussian distribution, and this latent variable is then used to reconstruct the input datapoint via the decoder. If a new latent variable drawn from the prescribed distribution is provided to the decoder, a new datapoint would be generated.

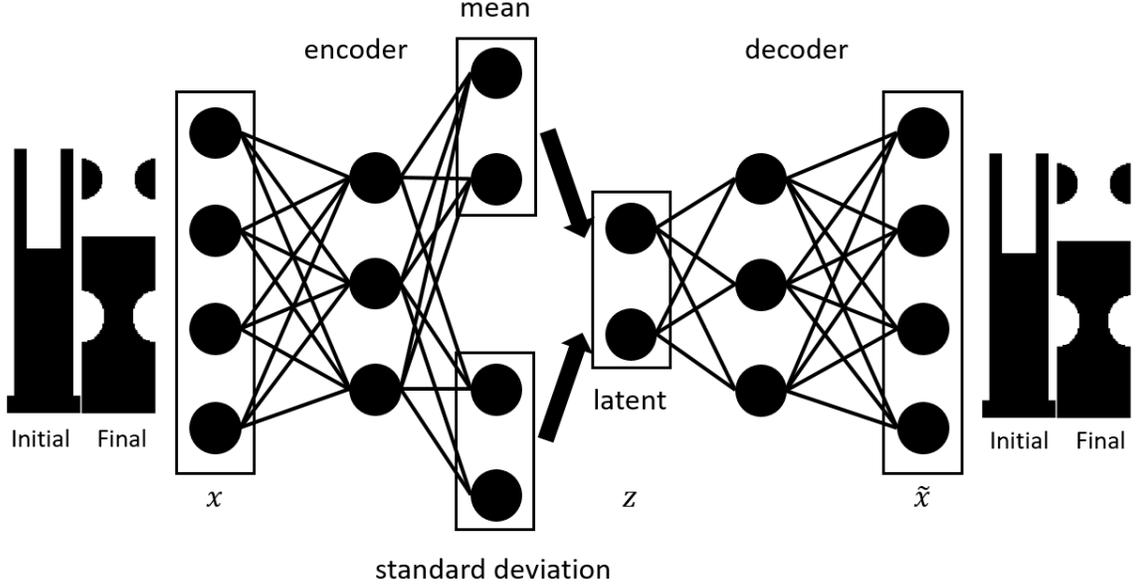

**Fig. 1.** The architecture of the proposed Variational Auto-encoder for inverse problems. The initial shape and final shape are put together to form one input datapoint. Both the encoder and the decoder consist of four layers with 512 nodes in each layer. The latent dimension is 100

To utilize VAE for inverse design, we pair the initial layout with the outcome to form one datapoint. In both our model problems, the desired outcome is also a shape. Hence our input datapoint is an image with the left part being the initial layout and the right part being the final shape as shown in **Fig. 1**. We first train the VAE to reconstruct input images. This is done by minimizing the loss function typically used in the VAE model [31]. This loss function is composed of the reconstruction error and the Kullback-Leibler (KL)-divergence error,

$$\mathcal{L} = \mathcal{L}_{reconst}(\tilde{x}, x) + \mathcal{L}_{KL}(z, unit\_gaussian) \tag{1}$$

where $\mathcal{L}_{reconst}(\tilde{x}, x) = binary\_corss\_entropy(\tilde{x}, x)$ denotes the reconstruction error and $\mathcal{L}_{KL}(z, unit\_gaussian)$ denotes the KL-divergence error, which measures the difference between the distribution of the latent variables and a normal distribution. We use ADAM optimizer [39] with default parameters. The learning rate is $1 \times 10^{-4}$ and the batch size is 128. Four layers with each layer consisting of 512 nodes are used in both the encoder and the decoder parts. The latent dimension is set to be 100. Activation function for the encoder and the decoder is exponential linear unit, and sigmoid function is chosen to be the activation function of the last layer, that is, the output layer.

After training, the decoder of the VAE model is used to perform the design. The goal of the inverse design is to find the initial shape with certain constraints for a targeted final structure. With the decoder, the design problem becomes finding an optimal latent variable that generates the targeted final structure. The following optimization problem is formulated for this purpose:

$$\hat{z} = \arg\min_{z}(\mathcal{L}(z) + \mathcal{R}(z)) \tag{2}$$

$$\mathcal{L}(z) = \|M \odot G(z) - y\|^2 \tag{3}$$

$$\mathcal{R}(z) = \alpha \|vol((1-M) \odot G(z)) - vol(y)\|^2 + \beta \cdot total\_variation(G(z)) \tag{4}$$

where $y$ represents the targeted final shape, $G(z)$ is the image generated by the VAE, $M \odot G(z)$ and $(1-M) \odot G(z)$ denote the final structure and the initial layout extracted form the generated image respectively. $\mathcal{R}(z)$ is a regularization term used to improve the quantity of the generated shapes. The first term in $\mathcal{R}(z)$ is to guarantee the volume denoted as $vol(\cdot)$ is conserved, which is needed in the first example. The second term is to guarantee the smoothness of the generated shapes. $total\_varaition$ is a measurement of noise, defining as the sum of the absolute differences in the values of neighboring pixels. $\alpha, \beta$ are two positive constants. In the first example, the two constants in $\mathcal{R}(z)$ are set as $\alpha = 0.1$ and $\beta = 0.2$. L-BFGS-B optimization [40] is used to solve Eq.(2). After the optimal latent variable $\hat{z}$ is obtained, the optimal initial shape can be extracted from the image $G(\hat{z})$ as $\hat{x} = (1-M) \odot G(\hat{z})$.

## 3. Results

*3.1. First example: inverse design for surface diffusion induced morphology change*

Surface diffusion induced morphology change refers to the shape and/or topology change in structures due to atom migration driven by chemical potential gradients along the surface. Denote by $\mu(\kappa)$ the increase in chemical potential per atom that is transferred from a point of zero curvature to a point of curvature $\kappa$ on the surface. It can be expressed as [41]

$$\mu(\kappa) = \kappa \gamma \Omega \tag{5}$$

where $\kappa$ is the surface mean curvature, $\gamma$ is the surface tension coefficient and $\Omega$ is the molar volume. Nonzero gradients of this chemical potential along the surface induce a drift of atoms with a surface flux given by

$$\boldsymbol{j}_s = -\frac{D_s \gamma \Omega \delta}{kT} \nabla_s \kappa \tag{6}$$

where $D_s$ is the surface diffusion coefficient, $\delta$ is the physical thickness of the diffusion layer, $k$ is Boltzmann constant, $T$ is temperature and $\nabla_s$ is the surface gradient. This surface flux results in a movement of surface in its normal direction with the velocity determined by the conservation law as,

$$\boldsymbol{v}_s = \left(\frac{D_s \gamma \Omega \delta}{kT} \Delta_s \kappa\right) \boldsymbol{n} = (C \Delta_s \kappa) \boldsymbol{n} \tag{7}$$

where $\Delta_s = \nabla_s \cdot \nabla_s$ is the Laplace-Beltrami operator, $\boldsymbol{n}$ is the outward normal vector at the surface and $C = \frac{D_s \gamma \Omega \delta}{kT}$. A typical example of morphology change due to surface diffusion is illustrated in **Fig. 2** in which the evolution of an initial trench structure shown in (a) due to surface diffusion is depicted. It can be seen from this figure that not only the shape but also the topology of the structure has changed with time.

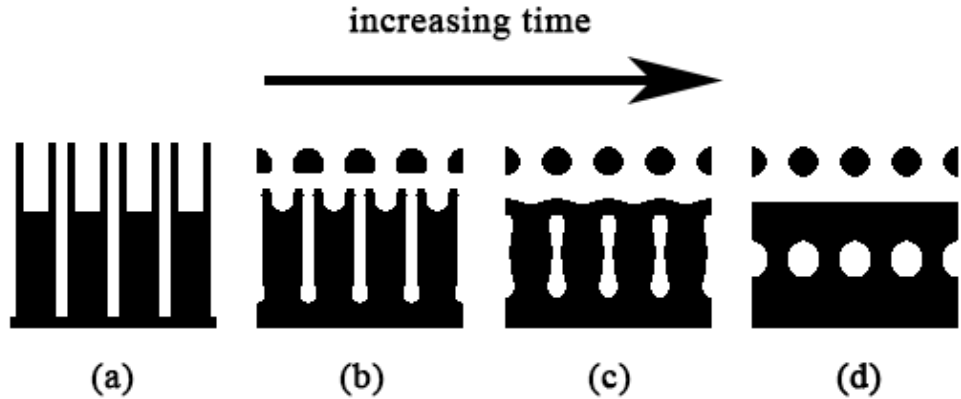

**Fig. 2.** A typical case of surface diffusion induced morphology change: (a) the initial structure (b) – (d) evolved structure at different time instants

Morphology change by surface diffusion can be found in many processes, for example, structural evolution of nanoporous metals during thermal coarsening and dealloying [42]. It affects material properties and performance and thus should be controlled if possible. Recently an innovative usage of this mechanism in microfabrication was proposed. It has been shown that buried cavities/microchannels can be self-assembled simply by annealing a prestructured silicon wafer at high temperature [43,44], without masks and bonding process. This technique also allows monolithic integration of MEMS-COMS [44], thus avoiding the material- and process-incompatibility issues inherent in the traditional integration schemes. Multiple microchannels with complicated architecture have been fabricated [45], and subattogram mass sensing and an active-matrix tactile sensor have also been demonstrated based on this fabrication technique [43,44].

The final stable structure after annealing is solely determined by the initial structure. In many applications, the architecture, the locations and the sizes of the cavities or bubbles have to be precisely controlled in order to achieve certain functionality. This calls for a careful design of the initial structure, which after the annealing process produces a final structure with the desired geometry. Such an inverse design problem is challenging due to the following reasons. Firstly structural evolution is a time-dependent problem and it is extremely time consuming to simulate the evolution process because, as it is well known, the surface diffusion is a numerically stiff problem. Calculating sensitivity information required in the gradient-based optimization methods would be even harder, if not impossible. Secondly the solution is non-unique in the sense that not only the initial structure is non-unique for the desired final structure, each intermediate structure during evolution is also a potential solution. However due to the fabrication constraints, only certain shapes of the initial structure are practically feasible. Thus this design problem would be a challenging one for conventional TO methods. Here we apply the proposed VAE based method to solve this inverse problem. We limit the shapes of the initial structure to be periodic trench structures, which are easy shapes to fabricate using conventional microfabrication techniques.

To train the VAE network, we create 10800 datapoints with 10000 datapoints for training and 800 for testing. Each datapoint contains a unit cell of the initial structure and a corresponding final structure obtained from an in-house surface diffusion solver described in the section of methodology. Both structures are represented by pixels with a size of $64 \times 256$, of which the periodicity 64 is in the $x$ direction. The initial structure is generated by placing one or two trenches randomly within the unit cell. The dimensions of each trench are chosen randomly as well. Samples of the training data are shown in **Fig. 3**. **Fig. 4** shows some samples of the datapoints randomly selected from the testing dataset and the reconstructed images, which are identical to the corresponding input images.

The trained VAE net is then used to design the initial structure for a targeted final structure. **Fig. 5** shows fifteen samples of the predicted initial structures for various targeted final structures drawn from the testing dataset. It can be seen that all the predicted structures are trench-like structures, indicating that the shape constraints have been leant very well by the constructed VAE model. To evaluate the design accuracy, we first smooth the designs by running a solver of the Cahn-Hillard equation with constant mobility for several iterations using a very small time step. We then run the surface-diffusion solver on the smoothed designs to obtain the corresponding final shapes. As shown in **Fig. 5**, the simulated final structures match quite well with the targeted final structures, demonstrating the success of the proposed design method. It also shows that the constructed VAE net has the ability to learn both the design constraints and the diffusion mechanism that evolves the initial shape to its final shape. To quantitatively measure the design accuracy, we use a binary accuracy defined as the percentage of the matched pixels, that is, the ratio of the total number of the corrected predicted pixels and the total number of pixels. The accuracy between the targeted final shapes and the simulated final shapes evaluated on all 800 testing data is 97.6%.

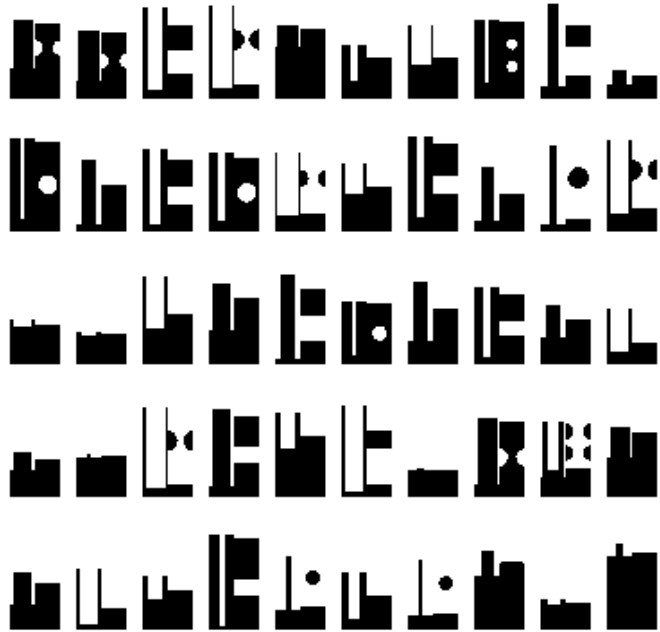

**Fig. 3.** Samples of the training data from the dataset

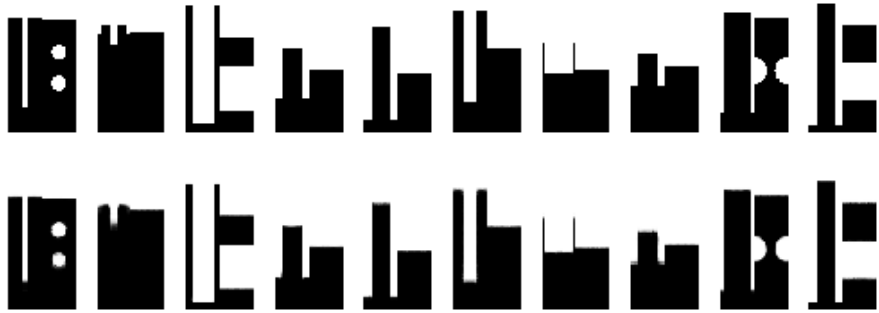

**Fig. 4.** Images selected from the testing data (first row) and the corresponding reconstructed images (second row)

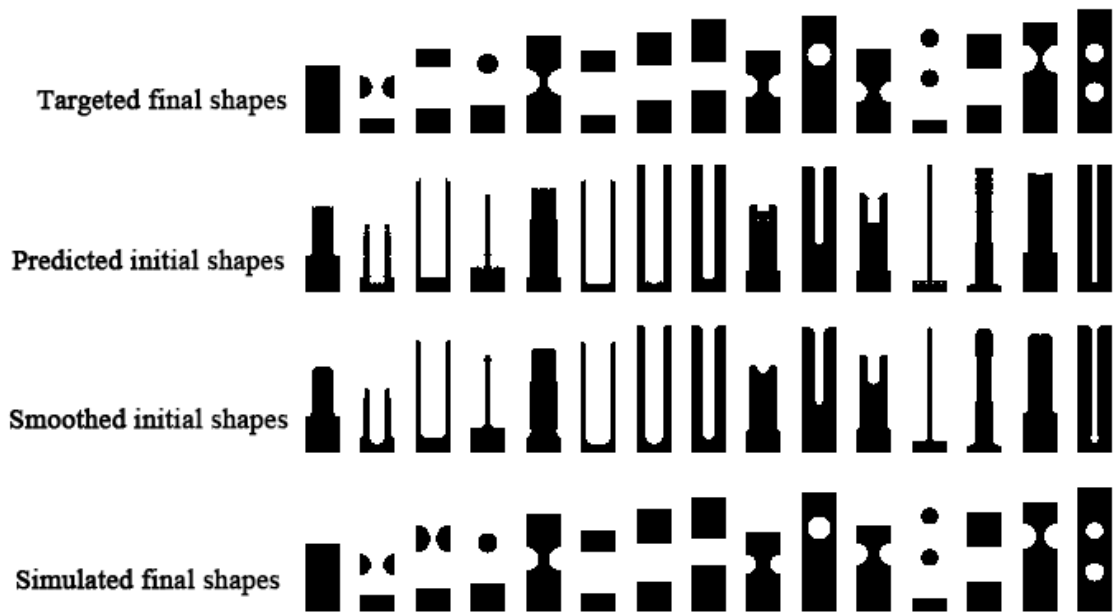

**Fig. 5.** Samples of the design results

*3.2. Second example: inverse mask synthesis*

Optical lithography is a standard micro/nano manufacturing process for fabricating integrated circuits (IC), microelectromechanical systems and other micro/nano devices/systems. The resolution of the photolithgrapy is hindered by the distortion in the transferred patterns on silicon wafers caused by the diffraction nature of the lights and the inherent limitations of the optical system, which results in short circuits, yield loss and malfunction of MEMS devices. One effecive way to correct the distortion is to per-compensate it by designing smart masks, that is, the inverse image synthesis method.

Much work has been done in this area and the majority employs methods similar to density-based TO methods [11,46]. Here we show that the proposed VAE based design method can be easily applied to synthesize smart masks. For demonstration purpose, we will consider a simple case similar to the one studied in [11]. As shown in Fig. 6, a pair of squares is to be printed onto the silicon wafer. Due to the distortion, which is modeled approximated using the convolution of the input pattern with a 2-D Gaussian kernel to calculate the aerial image formation and a hard-thresholding operation to simulate the resist effect [11], the transferred pattern is far from the desired squares. The goal of the inverse mask synthesis is to design a mask such that the transferred pattern is the two squares shown in **Fig. 6** (a).

Similar to the previous example, we first train the VAE model to reconstruct the input images, which compose of pairs of images: the mask and the corresponding transferred pattern. The architecture of the network is the same as that of the previous example, except that the latent dimension is chosen to be 10, and $\alpha$ and $\beta$ are set to be zero in this example. To generate the training data, a systematic method for producing mask images is developed based on the following strategies:

1. All masks should be symmetric since the desired transferred pattern is symmetric.
2. Masks should be as simple as possible. Here we will only consider images that are created by adding or removing rectangular shapes from the two squares at some control points as shown in **Fig. 7**. We start with a minimum number of control points and continue to add on more points if necessary.
3. The aspect ratio of the rectangular shapes is chosen to be within [0.5, 2.0]. This is to ensure that the synthesized mask is within a specified region.

According to the above criteria, a total of 10000 images are generated. **Fig. 8** shows some samples of the training images. To evaluate the generation capability of the VAE net, new images are generated from the trained network by sampling the latent variables. **Fig. 9** shows some of the generated new shapes. As it can be seen, all these new images satisfy the criteria specified above, demonstrating the learning capability of the network. We then perform the design procedure by searching a specific latent variable that produces a transferred shape to be as close as the two squares shown in **Fig. 6** (a). **Fig. 10** (a) shows the mask obtained from the design procedure, which is very different from the two squares and hard to obtain based on intuition. It is nevertheless still simple and generally satisfies the prescribed criteria. **Fig. 10** (b) shows the transferred pattern, which is much closer to the targeted shape than the one shown in **Fig. 6** (b).

(a) (b)

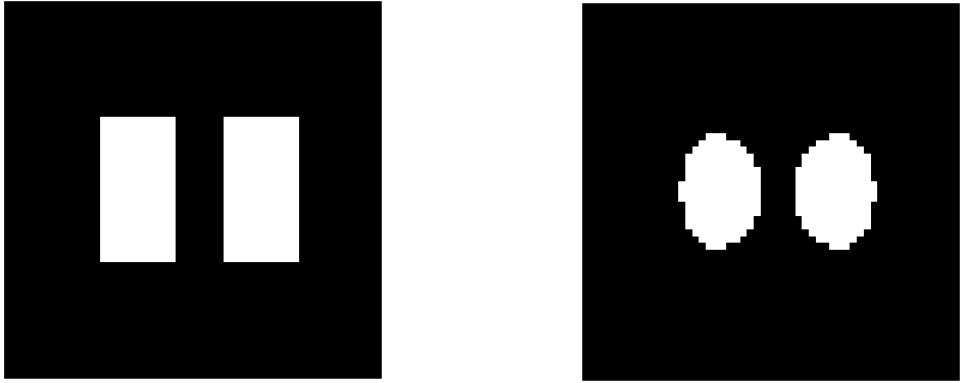

**Fig. 6.** (a) The initial mask; (b) The transferred pattern corresponding to the initial mask

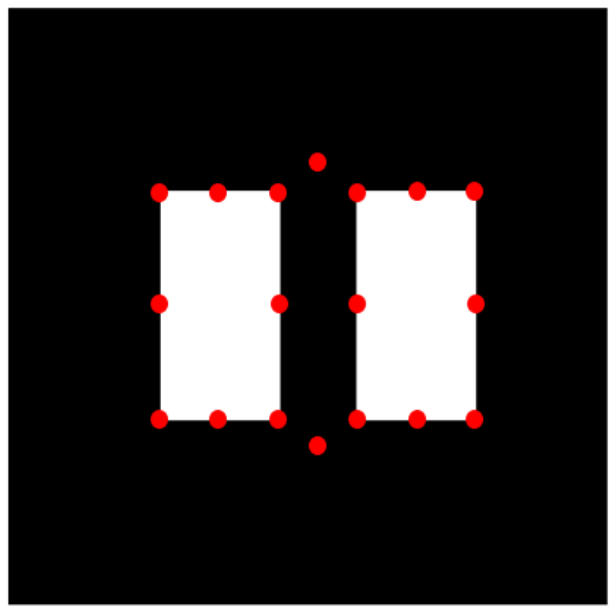

**Fig. 7.** Control points (red solid circles) for generating the training images

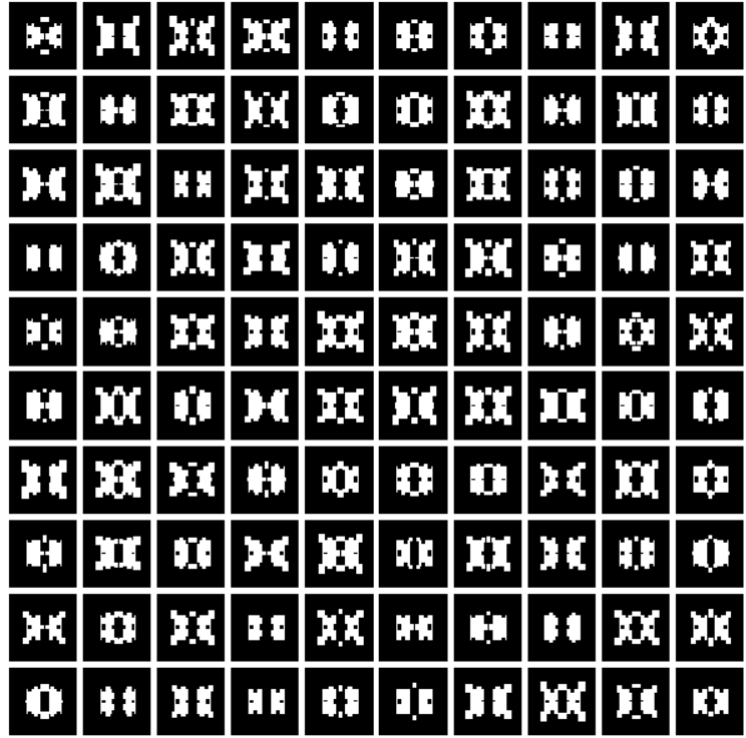

**Fig. 8.** Samples of the training mask images

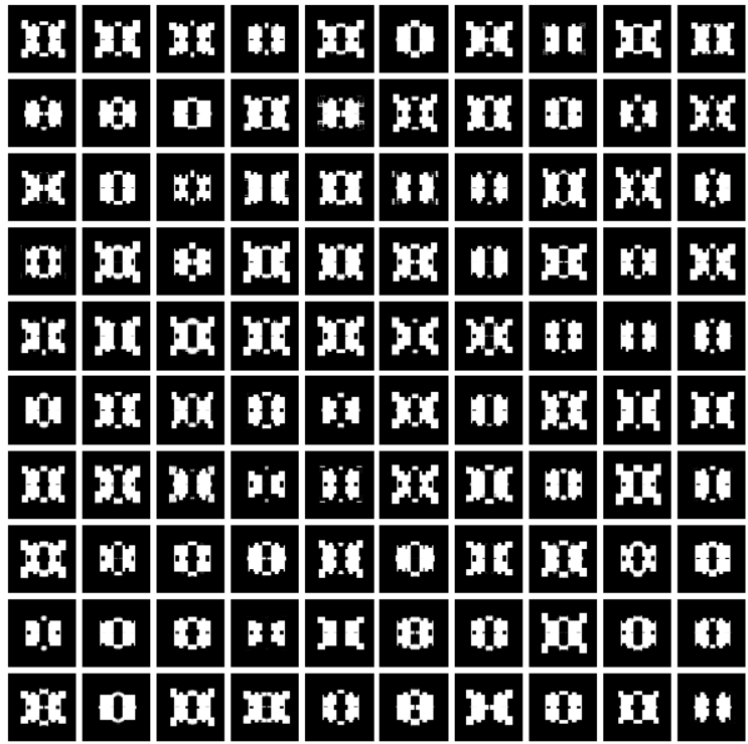

**Fig. 9.** Generated images by the VAE through sampling the latent variables

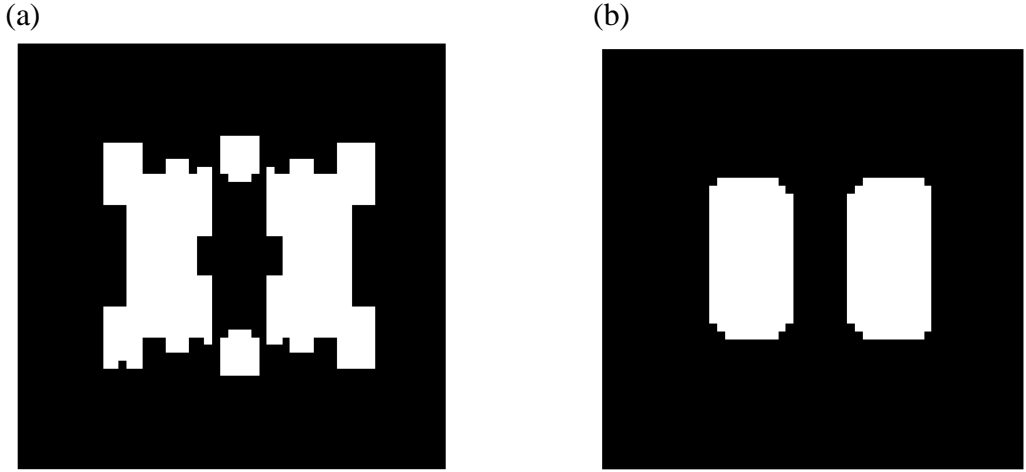

**Fig. 10.** (a) The predicted input mask; (b) The output shape of the input mask shown in (a)

## 4. Discussion

The proposed VAE based design method utilizes the learning capability of the VAE model to learn the hidden features shared between input data and the physical correlation between the inputs and the outputs, and uses the generative capability of the VAE to generate new design candidates that have similar features as those in the input data. As such, design constraints that are difficult to be formulated or incorporated into the conventional topological optimization procedures are automatically included in the design. As demonstrated in the two examples, the proposed VAE based inverse design method is very easy to implement and works for different physical problems despite the rather shallow network used. It does not need any sensitivity calculations. Only forward solutions are needed to generate the training data, which can be obtained either by modeling and simulation or from experimental measurements. This feature is particularly attractive for problems that are too complex to analyze. Moreover, the performance of the VAE model can always be enhanced by providing more training data and constructing a more advanced network utilizing deep network architecture and convolutional neural network. For example, in the case of mask synthesis, while the demonstrated VAE network is constructed specifically for the two-square case, a deeper VAE net can be constructed by providing more training data containing various types of mask images. Such a VAE net can then be used to design smart masks for any patterns, and it would be much more efficient than the conventional methods in which the design procedure, which includes numerous forward calculations and/or sensitivity analysis, must be re-run each time when a new mask is to be synthesized.

While in this paper only two design problems are solved to demonstrate the performance of the VAE design method, the application scope of the method is not limited just within the areas of these two problems. Any layout designs particularly those with complex constraints, for example, design of thin-film morphology for structural color applications [2], template design for directed assembly of block-copolymer morphologies [2] and design of heater surface to produce desired temperature and heat flux distribution [47,48], can be potential applications for the VAE method. The extension to those problems is, in principle, straightforward.

## Appendix: Surface diffusion solver

A phase-field model is used to describe interface evolution caused by surface diffusion, which is shown as follows:

$$\frac{\partial \phi}{\partial t} = \nabla \cdot \frac{9}{4\varepsilon} M(\phi) \nabla \mu \tag{A.1}$$

$$\mu = -\varepsilon \Delta \phi + f'(\phi) \tag{A.2}$$

$$f(\phi) = \frac{1}{4}(1 - \phi^2)^2 \tag{A.3}$$

$$M(\phi) = (1 - \phi^2)^2 \tag{A.4}$$

In these equations, $\phi$ denotes the order parameter for the phases of the system: $\phi = -1$ represents a void phase and $\phi = 1$ represents a solid phase; $\mu$ is the chemical potential; $\varepsilon$ is a parameter controlling the thickness of the interface; $f(\phi)$ is the bulk free energy; and $M(\phi)$ is the mobility. As $\varepsilon \to 0$ this model converges to Mullins sharp interface model (Eq. (7)) for surface diffusion, see [49].

The phase-field equations are solved numerically on a periodic domain by using an operator-splitting-based, quasi-spectral, semi-implicit time-stepping scheme. The semi-implicit scheme for Eq. (A.1) can be written in the following form [50],

$$\frac{\phi^{n+1} - \phi^n}{\Delta t} = \nabla \cdot \frac{9}{4\varepsilon} M(\phi) \nabla \mu|_n - B\Delta^2(\phi^{n+1} - \phi^n) + S\Delta(\phi^{n+1} - \phi^n) \tag{A.5}$$

where $B$ and $S$ are two positive stabilizing parameters chosen to guarantee the numerical stability. Performing Fourier transform on Eq. (A.5), we obtain,

$$\frac{\hat{\phi}^{n+1} - \hat{\phi}^n}{\Delta t} = F\left\{\nabla \cdot \frac{9}{4\varepsilon} M(\phi) \nabla \mu|_n\right\} - B(|k|^2)^2(\hat{\phi}^{n+1} - \hat{\phi}^n) + S|k|^2(\hat{\phi}^{n+1} - \hat{\phi}^n) \tag{A.6}$$

where $\hat{\phi}(k)$ is the Fourier transform of $\phi(x)$, and $F\{\cdot\}$ stands for Fourier transform.

Then, we can derive the following explicit time-stepping scheme in the spectrum space,

$$\hat{\phi}^{n+1} = \hat{\phi}^n + \frac{1}{1 + (S|k|^2 + B(|k|^2)^2)\Delta t}\left(F\left\{\nabla \cdot \frac{9}{4\varepsilon} M(\phi) \nabla \mu|_n\right\} + B(|k|^2)^2 \hat{\phi}^n - S|k|^2 \hat{\phi}^n\right) \tag{A.7}$$